\documentclass[12pt,a4paper]{article}

\usepackage{jheppub}
\usepackage{graphicx,amssymb,amsmath,amsfonts}
\usepackage{amsthm}
\usepackage[english]{babel}
\usepackage{hyperref}
\usepackage{cleveref}

\usepackage{braket}
\usepackage{enumerate}
\usepackage{bm}
\usepackage{fancyhdr}
\usepackage{calc}
\usepackage{float}
\usepackage{cancel}
\usepackage{subcaption}

\newcommand{\mn}{{\mu\nu}}
\newcommand{\rs}{{\rho\sigma}}
\newcommand{\mnrs}{{\mu\nu\rho\sigma}}

\newcommand{\V}{\mathbf}

\newcommand{\p}{{\partial}}

\newcommand{\zb}{{\bar{z}}}
\newcommand{\wb}{{\bar{w}}}
\newcommand{\gzz}{{\gamma_{z\bar{z}}}}
\newcommand{\guzz}{{\gamma^{z\bar{z}}}}

\newcommand{\gww}{{\gamma_{w\bar{w}}}}

\newcommand{\mJ}{\mathcal{J}}

\newcommand{\mM}{\mathcal{M}}

\newcommand{\mS}{\mathcal{S}}

\newcommand{\amp}{\braket{\text{out}|\mS|\text{in}}}
\newcommand{\td}[1]{\widetilde{d^3 #1}\,}

\newcommand{\tin}{{\text{in}}}
\newcommand{\tout}{{\text{out}}}
\newcommand{\fin}{{\Psi_\text{in}}}
\newcommand{\fout}{{\Psi_\text{out}}}

\newcommand{\wk}{{\omega_\V{k}}}

\newcommand{\WW}{{W^\dagger(\beta')W(\beta)}}

\makeatother

\title{BMS Supertranslation Symmetry Implies Faddeev-Kulish Amplitudes}
\author{Sangmin Choi, Ratindranath Akhoury}

\affiliation{Leinweber Center for Theoretical Physics, \\
Randall Laboratory of Physics, Department of Physics,\\
University of Michigan, Ann Arbor, MI 48109, USA}

\emailAdd{sangminc@umich.edu, akhoury@umich.edu}

\abstract{We show explicitly that, among the scattering amplitudes constructed from eigenstates of the BMS supertranslation charge, the ones that conserve this charge, are equal to those constructed from Faddeev-Kulish states. Thus, Faddeev-Kulish states naturally arise as a consequence of the asymptotic symmetries of perturbative gravity and all charge conserving amplitudes are infrared finite. In the process we show an important feature of the Faddeev-Kulish clouds dressing the external hard particles: these clouds can be moved from the incoming states to the outgoing ones, and vice-versa, without changing the infrared finiteness properties of S matrix elements. We also apply our discussion to the problem of the decoherence of momentum configurations of hard particles due to soft boson effects.

}

\begin{document}
{
\noindent

\maketitle
}

\section{Introduction}

Scattering amplitudes in gauge and gravitational theories suffer from infrared divergences, which (upon resummation) have
	the net effect of rendering all such amplitudes zero.
The traditional way of dealing with this problem is to employ the Bloch-Nordsieck
	method \cite{BN}, where one constructs an inclusive cross section out of all processes that are physically
	indistinguishable, i.e., including the contributions of undetectable soft bosons (photons or gravitions).
While this approach allows one to obtain cross sections that can be used to match with experiments,
	it has the shortcoming of giving up on the notion of a well-defined S-matrix element.
An alternative to this method is to use the asymptotic states of Faddeev and Kulish \cite{Kulish:1970ut}
	in place of Fock states.
These states can be interpreted as Fock states dressed by an infinite number of soft bosons,
	which are commonly referred to as the boson clouds.
It has been shown that using a set of such states as basis yields well-defined, infrared finite S-matrix elements.
	\cite{Chung:1965zza,Ware:2013zja}.

There has been work done in the recent years, for example 
	\cite{Kapec:2015ena,He:2014cra,He:2014laa,Strominger:2013jfa},
	that revealed the existence of an infinite number of degenerate vacua
	due to the spontaneous breaking of the asymptotic symmetries in gauge and gravity theories.
It has been argued that the vanishing of all S-matrix elements in the traditional approach is
	reflecting the fact that scattering processes induce a transition between the degenerate vacua,
	in a way that conserves the charges of the broken symmetries.
This paved the way to understanding the connection of asymptotic symmetries to
	the formalism of Faddeev and Kulish.
The boson clouds of the Faddeev-Kulish states have been shown,
	in \cite{Gabai:2016kuf} for QED and \cite{Choi:2017bna} for gravity,
	to precisely cancel the vacuum transitions induced by the scattering operator,
	which explains why Faddeev-Kulish states yield well-defined S-matrix elements.
Also, it has been shown in \cite{Kapec:2017tkm} that a gauge-invariant formulation of the
	charged particles in QED analogous to \cite{Dirac,Bagan:1999jf} yields
	coherent states that are essentially equivalent to Faddeev-Kulish states,
	which turn out to be the charge eigenstates of the large gauge symmetry.

In this paper, we take one step further and argue that the infrared-finite scattering amplitudes
	constructed using the Faddeev-Kulish states (henceforth referred to as Faddeev-Kulish amplitudes) 
	naturally arise as a consequence of asymptotic symmetry. Thus, since Faddeev-Kulish amplitudes are infrared finite, so are
	all BMS supertranslation charge conserving amplitudes.
We demonstrate this in the case of perturbative quantum gravity, by constructing eigenstates
	of BMS supertranslation charge and showing that any scattering amplitude that conserves
	this charge is equivalent to the Faddeev-Kulish amplitude.
In this process, we show that the graviton clouds ``weakly commute" with the scattering operator,
	in the sense that clouds in the incoming state can freely be moved to the outgoing state and vice versa.
Our work provides a natural proof of the conjecture made in \cite{Kapec:2017tkm},
	which claims that amplitudes conserving charges of asymptotic symmetries are infrared finite.
We conclude with an application of our results to the study initiated in \cite{Carney:2017jut}, where information theoretic properties of low energy photons and gravitons are analyzed through the study of the relevant density matrices. Our approach here, in contrast to \cite{Carney:2017jut}, is to derive expressions for the density matrices which satisfy conservation of BMS supertranslation charge at all stages.
If the measurements are sensitive only to the momenta of the hard matter particles, then the conclusions of \cite{Carney:2017jut,Carney:2017oxp} are unchanged. There is a decoherence of the momentum configurations of these particles. This decoherence and the consequent high degree of correlations between the hard and soft quanta was also noted independently in \cite{Strominger:2017aeh} using a different approach.

The paper is organized as follows.
In section \ref{EIGENSTATE}, we construct the eigenstates of BMS supertranslation charge,
	and study the implications of the charge conservation on the scattering amplitudes.
We establish in section \ref{MOVECLOUD} that the Faddeev-Kulish graviton clouds weakly commute with the scattering operator.
This result is used in section \ref{SAME} to show the equality between
	amplitudes that conserve the supertranslation charge and the Faddeev-Kulish amplitudes.
In section \ref{DECOHERENCE}, we apply the preceding results to the analysis of \cite{Carney:2017jut}.
We wrap up with a brief discussion in section \ref{DISCUSSION}.

\section{BMS charge and eigenstates}\label{EIGENSTATE}

In order to establish notation and to make connections with earlier work, we will begin with a review of BMS symmetry and the conserved charges.
As is customary, we will employ the retarded coordinates $(u,r,z,\zb)$, defined in terms of the Cartesian coordinates
	$(t,x_1,x_2,x_3)$ as
\begin{align}
	u = t-r,\qquad r^2=x_1^2+x_2^2+x_3^2,\qquad z=\frac{x_1+ix_2}{r+x_3}.
\end{align}
Here $u$ is the retarded time and $z$ is the complex coordinate on the unit 2-sphere
	with the metric $\gzz = \frac{2}{(1+z\zb)^2}$.
Then in the Bondi gauge \cite{Bondi:1962px,Sachs:1962wk}, the asymptotically flat metric has
	the expansion \cite{Strominger:2013jfa}
\begin{align}
\begin{split}
	ds^2 =& -du^2 - 2dudr + 2r^2 \gzz dz d\zb
		\\ &
		+\frac{2m_B}{r}du^2 + rC_{zz}dz^2 + rC_{\zb\zb}d\zb^2 + D^zC_{zz}dudz + D^\zb C_{\zb\zb}dud\zb
		\\ & +\cdots,
\end{split}
\end{align}
where $m_B$ is the Bondi mass aspect and $D^z$, $D^\zb$ are the 2-sphere covariant derivatives.
The gravitational radiation is characterized by the Bondi news tensor $N_{zz}=\p_u C_{zz}$.

The BMS supertranslation charge for a 2-sphere function $f=f(w,\wb)$ is then 
\begin{align}
	Q(f) = Q_S(f) + Q_H(f),
\end{align}
where, explicit expressions for the soft part $Q_S$ and the hard part $Q_H$ are given in \cite{Strominger:2017zoo,Campiglia:2015kxa}.
We are interested in these expressions at the leading terms in the large-$r$ expansion
which are known to be gauge-invariant \cite{Avery:2015gxa}.

The action of the hard charge $Q_H$ on a Fock state of $N$ massive particles can be expressed as \cite{Campiglia:2015kxa}
\begin{align}
	Q_H\ket{\V{p}_1,\ldots,\V{p}_N}
		&= \sum_{i=1}^N \tilde{f}(p_i)\ket{\V{p}_1,\ldots,\V{p}_N},
\end{align}
where $p_i^\mu = (E_k,\V{p}_i)$, and
\begin{align}
	\tilde{f}(p)
		= -\frac{1}{2\pi}
			\int d^2w \,
			\frac{(\epsilon^+(w,\wb)\cdot p)^2}{p\cdot \hat{x}_w}
			D^2_\wb f(w,\wb).
\end{align}
Here $\hat{x}^\mu_w=(1,\hat{\V{x}}_w)$ with the unit vector $\hat{\V{x}}_w$ pointing in the direction $(w,\wb)$,
	and the polarization vectors have components
\begin{align}
	\epsilon^{-\mu}(z,\zb)=\frac{1}{\sqrt{2}}(z,1,i,-z)
	\quad\text{and}\quad
	\epsilon^{+\mu}(z,\zb)=\frac{1}{\sqrt{2}}(\zb,1,-i,-\zb).
\end{align}
The action of the soft charge $Q_S$ on the same state is \cite{Strominger:2013jfa}
\begin{align}
	Q_S\ket{\V{p}_1,\ldots,\V{p}_N} &=
	-\frac{1}{8\pi G}\int du\, d^2w\, \gww N^{\wb\wb}D_\wb^2 f
	\ket{\V{p}_1,\ldots,\V{p}_N}.
\end{align}
Conservation of BMS supertranslation charges imply,
\begin{align}\label{Scons}
	\braket{\tout|\, [Q(f),\mS]\,|\tin} = 0,
\end{align}
which should hold for all functions $f(w,\wb)$.
In particular, let us choose
\begin{align}
	f(w,\wb) = \frac{(1+w\wb)(\wb-\zb)}{(1+z\zb)(w-z)},
\end{align}
such that \cite{Campiglia:2015kxa}
\begin{align}
	D_\wb^2 f(w,\wb) = 2\pi \delta^2(w-z).
\end{align}
With this choice, the conservation law  \eqref{Scons} reads
\begin{align}\label{rawcons}
	\frac{\gzz}{4 G}\int_{-\infty}^\infty du\braket{\tout|(N^{\zb\zb}\mS - \mS N^{\zb\zb})|\tin}
	= -\sum_i \eta_i
		\frac{(p_i\cdot \epsilon^+(z,\zb))^2}{p_i\cdot \hat{x}_z}
		\braket{\tout|\mS|\tin},
\end{align}
where the sum on the RHS runs over all external particles and $\eta_i=+1$ ($-1$) if $i$ is an
	outgoing (incoming) particle.
Let us define the operator
\begin{align}
	N(z,\zb) \equiv \gzz \int^\infty_{-\infty} du\,N^{\zb\zb}
		= \guzz \int^\infty_{-\infty}du\,N_{zz}.
\end{align}
Then \eqref{rawcons} becomes
\begin{align}
	\braket{\tout|(N(z,\zb)\mS - \mS N(z,\zb))|\tin}
	= -\frac{\kappa^2}{8\pi}\sum_i \eta_i
		\frac{(p_i\cdot \epsilon^+(z,\zb))^2}{p_i\cdot \hat{x}_z}
		\braket{\tout|\mS|\tin},
\end{align}
where $\kappa = \sqrt{32\pi G}$.
If the in- and out-states are eigenstates of $N(z,\zb)$ such that
\begin{align}
	\bra{\tout}N(z,\zb) &= N_\tout \bra{\tout}
	\qquad\text{and}\qquad
	N(z,\zb)\ket{\tin} = N_\tin \ket{\tin},
\end{align}
then we obtain
\begin{align}\label{refined_cons}
	\left(N_\tout - N_\tin\right)\braket{\tout|\mS|\tin}
	= \Omega^\text{soft}
		\braket{\tout|\mS|\tin},
\end{align}
with a soft factor that is analogous to that of \cite{Kapec:2017tkm}:
\begin{align}
\Omega^\text{soft} = -\frac{\kappa^2}{8\pi}\sum_i \eta_i
		\frac{p_i^\mu p_i^\nu}{p_i\cdot \hat{x}_z}\epsilon^+_\mn.
\end{align}
To see what the eigenstates look like, we first note that $N(z,\zb)$ can be expressed
	in terms of the graviton creation and annihilation operators as \cite{He:2014laa}
\begin{align}
	N(z,\zb) = -\frac{\kappa}{8\pi}\lim_{\omega\to 0}
		\left[
			\omega a_+(\omega \hat{x}_z) + \omega a_-^\dagger (\omega \hat{x}_z)
		\right ].
\end{align}
This suggests that the eigenstate should take some form of a coherent graviton state.
Next, consider the following state
\begin{align}\label{Nstate}
	\ket{N}=\exp\left\{
		\int\td{k} N^\mn(k) \left[a^\dagger_\mn(k) - a_\mn(k)\right]
	\right\}\ket{0},
\end{align}
where $\td{k} = \frac{d^3k}{(2\pi)^3 (2\wk)}$ is the Lorentz-invariant measure,
\begin{align}
	a_\mn^\dagger(k) = \sum_r \epsilon^r_\mn(k) a^{r\dagger}(k),
	\qquad
	a_\mn(k) = \sum_r \epsilon^{r*}_\mn(k) a^{r}(k),
\end{align}
$N^{\mu\nu}$ is an arbitrary symmetric tensor and the sum runs over all polarizations, including the unphysical ones.
We will next show that if the symmetric tensor $N^\mn(k)$ has soft poles, then the above state is an eigenstate of both $\lim \omega a_+$ and $\lim\omega a^\dagger_-$. Indeed,
\begin{align}
	\lim_{\omega\to 0}\omega a_+(\omega \hat{x}_z)\ket{N}
	 &= \lim_{\omega\to 0}\omega
		\left[
			a_+(\omega \hat{x}_z), \int\td{k} N^\mn(k) \left(a^\dagger_\mn(k) - a_\mn(k)\right)
		\right]
		\ket{N}
	\\ &= \lim_{\omega\to 0}\frac{\omega}{2}
		N_\mn(\omega \hat{x}_z) I^\mnrs\epsilon^+_\rs(z,\zb)
		\ket{N}
	\\ &= \lim_{\omega\to 0}\omega
		N^\mn(\omega \hat{x}_z) \epsilon^+_\mn(z,\zb)
		\ket{N}.
\end{align}
Thus we see that the eigenvalue is non-zero only if $N^\mn$ has poles for soft momenta. Similarly,
\begin{align}\label{convac}
	\lim_{\omega\to 0}\omega a_-^\dagger(\omega \hat{x}_z)\ket{N}
		&= \lim_{\omega\to 0}\omega
		N^\mn(\omega \hat{x}_z) \epsilon^+_\mn(z,\zb)
		\ket{N}.
\end{align}
It should be noted that in \eqref{convac}, the term with the creation operator acting on the vacuum vanishes upon taking the soft
 limit $\omega \rightarrow 0$.
From this we can immediately see that $\ket{N}$ is an eigenstate of $N(z,\zb)$, i.e.,
\begin{align}
	N(z,\zb) \ket{N} = -\frac{\kappa}{4\pi}
		\left(\, \lim_{\omega\to 0}\omega N^\mn \epsilon^+_\mn\right)\ket{N}.
\end{align}
In particular, the Fock vacuum $\ket{0}$ , which corresponds to $N^\mn=0$, is itself an eigenstate with eigenvalue $0$.
Later, when considering S matrix elements, we will for convenience put $N^\mn=0$ for the incoming state, 
which amounts to assuming that the incoming state is a Fock state. This does not entail a loss of generality because
as can be seen from \eqref{refined_cons}, it is only the difference $N_\tout^\mn-N_\tin^\mn$ that matters.
Similarly, the bra state
\begin{align}
	\bra{N}=\bra{0}\exp\left[
		-\int\td{k} N^\mn \left(a^\dagger_\mn - a_\mn\right)
	\right]
\end{align}
is an eigenstate of $N(z,\zb)$:
\begin{align}
	\bra{N}N(z,\zb) = -\frac{\kappa}{4\pi}\bra{N}
		\left(\, \lim_{\omega\to 0}\omega N^\mn \epsilon^+_\mn\right).
\end{align}
We want to treat these eigenstates as alternative vacuums, so we will restrict the
	momentum integrals to run over only the soft momenta.
With these choices, $\ket{N}$ will remain an eigenstate with any number of hard particle operators acting on it.

The conservation law $N_\text{out}-N_\text{in}=\Omega^\text{soft}$ implied by \eqref{refined_cons} is then
\begin{align}
	\lim_{\omega\to 0}\omega \big[N^\mn_\text{out}(\omega\hat{x}_z)
			- N^\mn_\text{in}(\omega\hat{x}_z)\big] \epsilon^+_\mn(z,\zb)
		&= \frac{\kappa}{2}\sum_i \eta_i\frac{p_i^\mu p_i^\nu }{p_i\cdot \hat{x}_z}\epsilon^+_\mn(z,\zb).
\end{align}
As shown above, the leading soft terms in $N^{\mu\nu}$ are the only ones contributing to the eigenvalue, which therefore satisfy
\begin{align}\label{eigen_cons}
	N^\mn_\text{out}(k) - N^\mn_\text{in}(k)
		= \frac{\kappa}{2}\sum_i \eta_i\frac{p_i^\mu p_i^\nu}{p_i\cdot k},
\end{align}
where we have put $k = \omega \hat{x}_z$.
We should emphasize that either this conservation law is satisfied or the amplitude $\amp$ vanishes.
This implies that if the initial state is built on the Fock vacuum $\ket{0}$, i.e.
\begin{align}
	\ket{\text{in}} = \prod_{i\in\text{in}}b^\dagger(p_i)\ket{0},
\end{align}
where $b^\dagger$ is the creation operator of hard massive particles, then
	this state does not scatter into any state built on the same vacuum $\ket{0}$,
	since in that case $N_\text{out}=N_\text{in}=0$, thereby violating the conservation law
	$N_\tout - N_\tin = \Omega^\text{soft}$.
Instead, scattering must take place into states built on the vacuum $\ket{N_\tout}$ with
\begin{align}
	N^\mn_\text{out}=\frac{\kappa}{2}\sum_i \eta_i\frac{p_i^\mu p_i^\nu}{p_i\cdot k}.
\end{align}
Such states therefore have the form, (see Eq. \eqref{Nstate})
\begin{align}
	\bra{\text{out}}=\bra{0}\left[\prod_{j\in\text{out}}b(p_j)\right]
	\exp\left[
		-\frac{\kappa}{2}\sum_i\eta_i \int\td{k} \frac{p_i^\mu p_i^\nu}{p_i\cdot k}
		(a^\dagger_\mn - a_\mn)
	\right].
\end{align}
The scattering amplitude now can be written in the form:
\begin{align}\label{amplitude}
	\bra{\text{out}}
	\mS
	\ket{\text{in}}
	= \bra{\fout}
		\exp\left[
			-\frac{\kappa}{2}\sum_i\eta_i\int\td{k} \frac{p_i^\mu p_i^\nu}{p_i\cdot k}
			(a^\dagger_\mn - a_\mn)
		\right]
		\mS
		\ket{\fin},
\end{align}
where $\fout$, $\fin$ denote the usual Fock states for the hard particles.
The form of \eqref{amplitude} is reminiscent of the Faddeev-Kulish amplitudes.
In the following two sections we will spell out this equivalence more precisely.
It will turn out that any amplitude that obeys the conservation law \eqref{eigen_cons},
an example being \eqref{amplitude}, is equal to the Faddeev-Kulish amplitude and is therefore IR-finite.

\section{Relation to Faddeev-Kulish amplitudes} 

As a first step in establishing this equality, we will demonstrate a crucial feature of the Faddeev-Kulish amplitudes
which, although technical, has important physical consequences.
Since a Faddeev-Kulish amplitude is constructed by dressing each external particle with its cloud of soft gravitons,
	an amplitude with $n$ incoming and $n'$ outgoing particles necessarily
	has $n$ clouds on the right of the scattering operator $\mS$, and $n'$ clouds on the left.
Although the clouds commute with each other, it was not clear how things change if, for example, one moves
	a cloud dressing an incoming particle (therefore sitting on the right of $\mS$) to the left of $\mS$. In this connection, 
based on the conservation of supertranslation charge and the crossing symmetry, the authors of \cite{Kapec:2017tkm}
	conjectured that such amplitudes exhibit the same cancellation of IR divergences. In this section, we will explicitly show that the 
	clouds ``weakly commute" with $\mS$, in the sense that in an $\mS$ matrix element, any incoming cloud
	can be moved to the outgoing state without affecting the amplitude, and vice versa.
This result proves the aforementioned conjecture, since it follows that the amplitudes considered
	in \cite{Kapec:2017tkm} are equal to the Faddeev-Kulish amplitude.
Then in the next section, we will use this to show that any amplitude that conserves supertranslation charge,
	for example \eqref{amplitude}, is equal to the Faddeev-Kulish amplitude with the same external particle configuration.
This will establish the notion that Faddeev-Kulish amplitudes naturally arise from the charge conservation of
asymptotic symmetries.

In order to relate \eqref{amplitude} to the Faddeev-Kulish amplitude,
let us denote its left hand side by $\mM$,
\begin{align}
	\mM &= \bra{\fout}
		\exp\left[
			-\frac{\kappa}{2}\sum_i\eta_i\int\td{k} \frac{p_i^\mu p_i^\nu}{p_i\cdot k}
			(a^\dagger_\mn - a_\mn)
		\right]
		\mS
		\ket{\fin}.
\end{align}
Next, consider another amplitude $\mM_c$, given by
\begin{align}
	\mM_c &= \bra{\fout}
		\exp\left\{
			-\frac{\kappa}{2}\sum_i\eta_i\int\td{k} \left[\frac{p_i^\mu p_i^\nu}{p_i\cdot k} + \frac{c^\mn(p_i,k)}{\wk}\right]
			(a^\dagger_\mn - a_\mn)
		\right\}
		\mS
		\ket{\fin}
	\\ &= \label{new_amp}
		\bra{\fout}\exp\left[-\sum_i \eta_i R_f(p_i)\right]\mS\ket{\fin},
\end{align}
where we inserted a term proportional to $c^\mn/\wk$ to the argument of the exponential.
Here $c^\mn(p,k)$ is the tensor of \cite{Ware:2013zja} that parametrizes the asymptotic space, and
\begin{align}\label{Rf}
	R_f(p_i) = \frac{\kappa}{2}\int \td{k} \left[\frac{p_i^\mu p_i^\nu}{p_i\cdot k} + \frac{c^\mn(p_i,k)}{\wk}\right]
		(a^\dagger_\mn - a_\mn)
\end{align}
is the anti-Hermitian operator appearing in the construction of Faddeev-Kulish state
	\cite{Ware:2013zja} with $\phi = 1$.
In contrast to $\mM$ and $\mM_c$, the IR-finite Faddeev-Kulish amplitude $\mM_\text{FK}$ is given by
\begin{align}\label{FKamp}
\mM_\text{FK} = \braket{\fout|\exp\left[-\sum_{i\in\tout} R_f(p_i)\right]
	\mS \exp\left[\sum_{i\in\tin} R_f(p_i)\right]|\fin}.
\end{align}
We aim to establish $\mM_{FK}=\mM_c=\mM$.

\subsection{Moving the graviton clouds}\label{MOVECLOUD}

Let us start by considering the simplest case, i.e.,  the Faddeev-Kulish amplitude for single-particle external states to leading order in the interaction.
We follow the shorthand notations used in \cite{Choi:2017bna}:
\begin{align}
P_\mn(p,k) = \frac{\kappa}{2}\left(\frac{p_\mu p_\nu}{p\cdot k}\right),
\qquad
C_\mn(p,k) = \frac{\kappa}{2}\frac{c_\mn(p,k)}{\wk},
\end{align}
and $S_\mn(p,k) = P_\mn(p,k)+C_\mn(p,k)$.
These allow us to write, (see \cite{Choi:2017bna} for details)
\begin{align}
	\mM_\text{FK} =
		\bra{0}
		b(p_f)
		e^{
			-S_f \cdot(a^\dagger - a)
		}
		\mS
		e^{
			S_i \cdot(a^\dagger - a)
		}
		b^\dagger(p_i)
		\ket{0},
\end{align}
where $S_f^\mn \equiv S^\mn(p_f,k)$ and $S_i^\mn \equiv S^\mn(p_i,k)$.
The subscript FK is written to emphasize that this is a Faddeev-Kulish amplitude.
In what follows we will employ the following notation,
\begin{align}
	S\cdot (a^\dagger-a) \equiv \int\td{k}S^\mn(a^\dagger_\mn - a_\mn),
\end{align}
and
\begin{align}
	S_f\cdot I \cdot S_i \equiv \int\td{k} S_f^\mn I_\mnrs S_i^\rs,
\end{align}
where
\begin{align}
I^\mnrs=\eta^{\mu\rho}\eta^{\nu\sigma}+\eta^{\mu\sigma}\eta^{\nu\rho}-\eta^{\mn}\eta^{\rs}.
\end{align}
Upto the one loop order, this amplitude is
\begin{align}
	\mM_\text{FK} =
		\bra{0}
		b(p_f)
		\left(
			1+S_f \cdot a - \frac{1}{4}S_f\cdot I\cdot S_f
		\right)
		\mS
		\left(
			1+S_i \cdot a^\dagger - \frac{1}{4}S_i \cdot I\cdot S_i
		\right)
		b^\dagger(p_i)
		\ket{0}.
\end{align}
Working out the infrared divergences (see \cite{Choi:2017bna} for details), we see that they factor out and cancel as
\begin{align}
	\Bigg(
		1
		-\underbrace{\frac{1}{4}P\cdot I \cdot P}_\text{virtual}
		+\underbrace{\frac{1}{2}S\cdot I \cdot P}_\text{interacting}
		-\underbrace{\frac{1}{4}S\cdot I \cdot S}_\text{cloud-to-cloud}
	\Bigg)\braket{p_f|\mS|p_i}
	= \braket{p_f|\mS|p_i},
\end{align}
where $P=P_f-P_i$ and $S=S_f-S_i$. Note that the various infrared divergent contributions are indicated in braces. These are (1) corrections due to virtual graviton exchange, (2) the interacting graviton corrections arising from gravitons connecting the Faddeev-Kulish clouds to external legs, and finally (2) corrections due to cloud-to-cloud graviton exchanges. These have been discussed in detail in appendix B of \cite{Choi:2017bna}.

\begin{figure}
	\centering
    \begin{subfigure}{0.24\textwidth}
		\includegraphics[width=\textwidth]{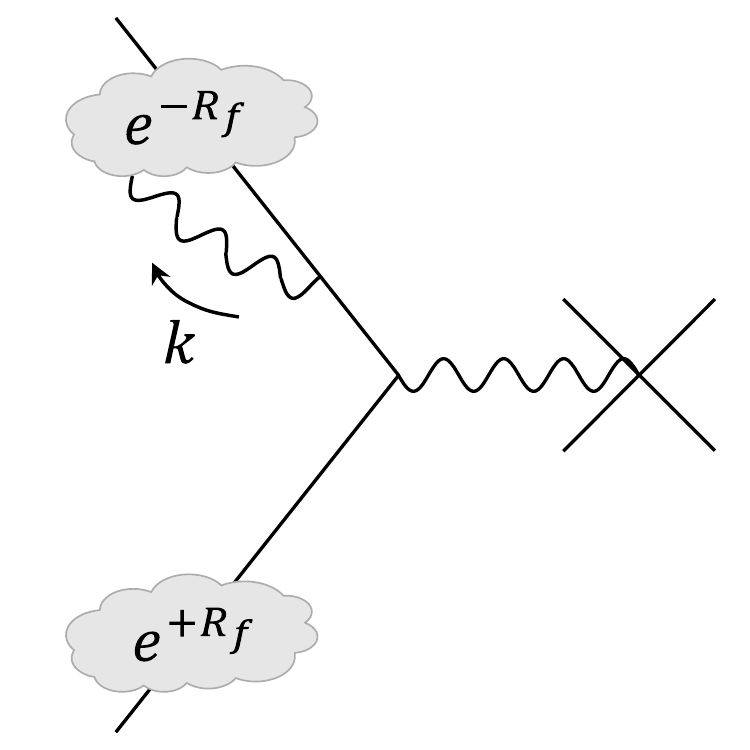}
		\caption{}
    \end{subfigure}
    \begin{subfigure}{0.24\textwidth}
		\includegraphics[width=\textwidth]{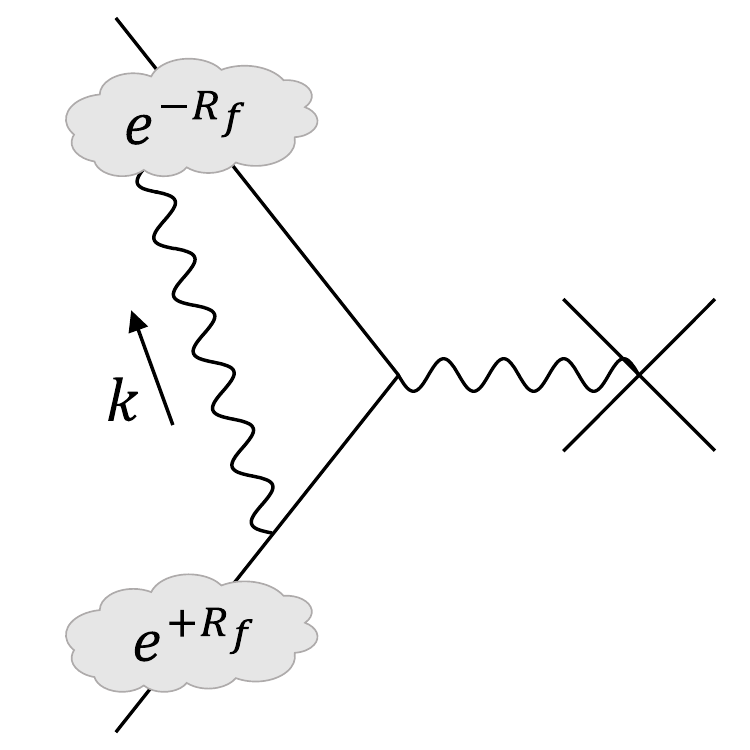}
		\caption{}
    \end{subfigure}
    \begin{subfigure}{0.24\textwidth}
		\includegraphics[width=\textwidth]{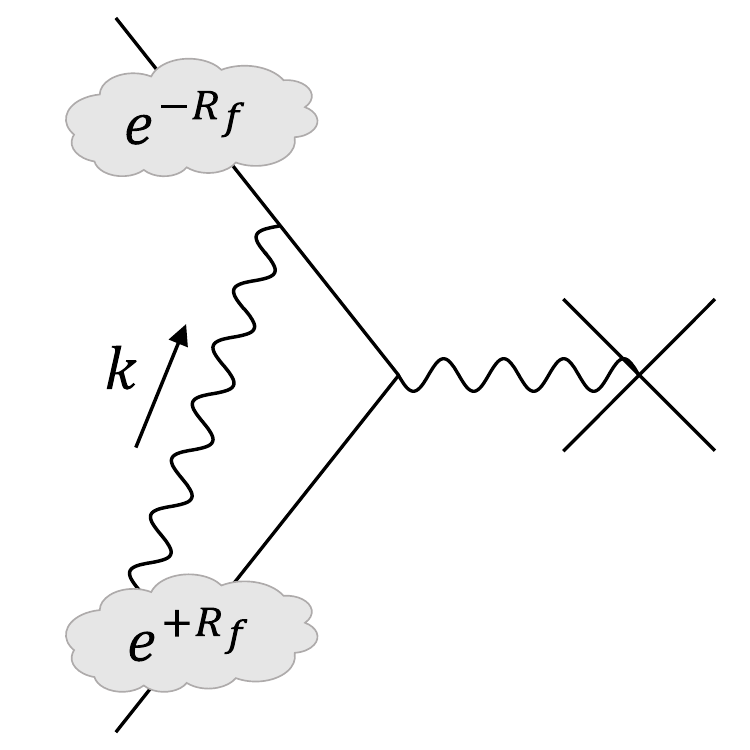}
		\caption{}
    \end{subfigure}
    \begin{subfigure}{0.24\textwidth}
		\includegraphics[width=\textwidth]{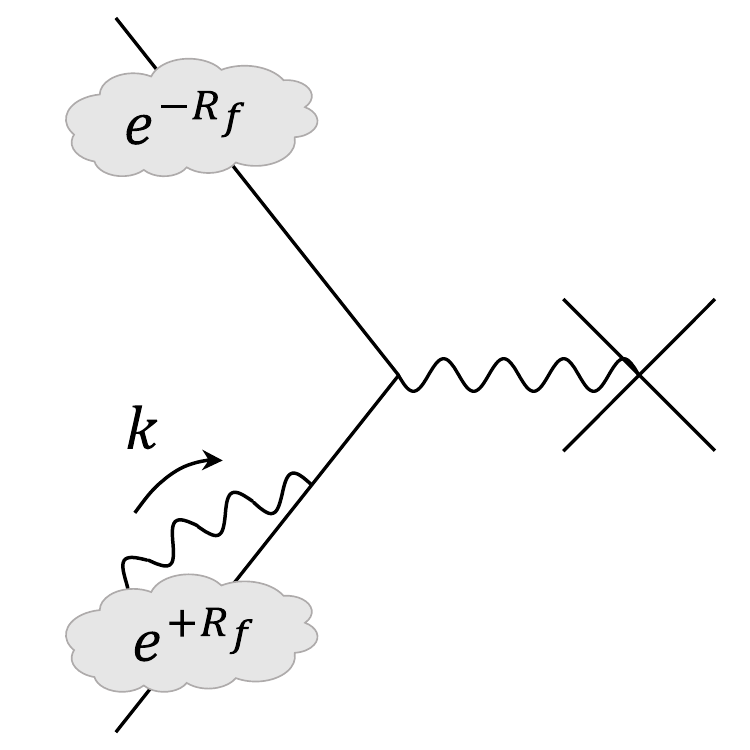}
		\caption{}
    \end{subfigure}
    \begin{subfigure}{0.24\textwidth}
		\includegraphics[width=\textwidth]{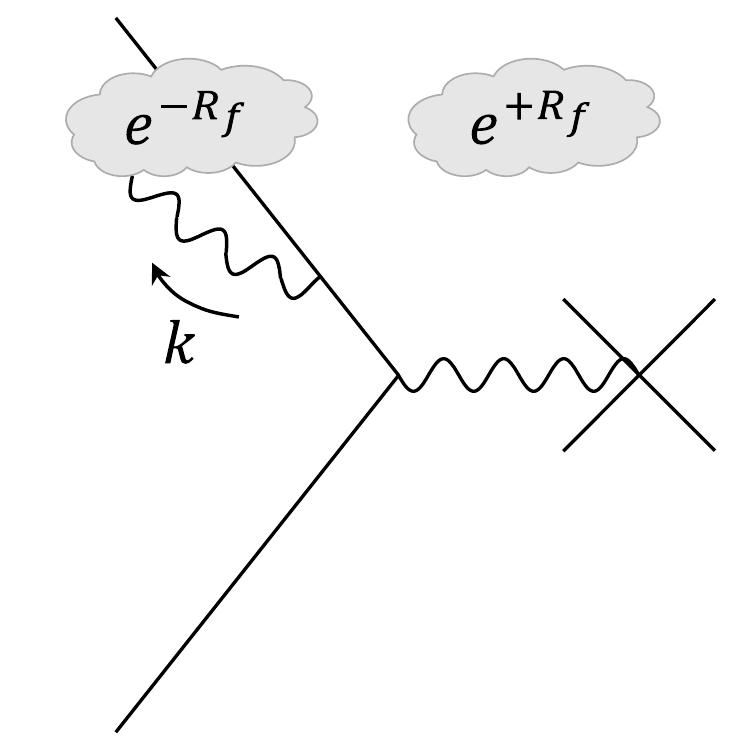}
		\caption{}
    \end{subfigure}
    \begin{subfigure}{0.24\textwidth}
		\includegraphics[width=\textwidth]{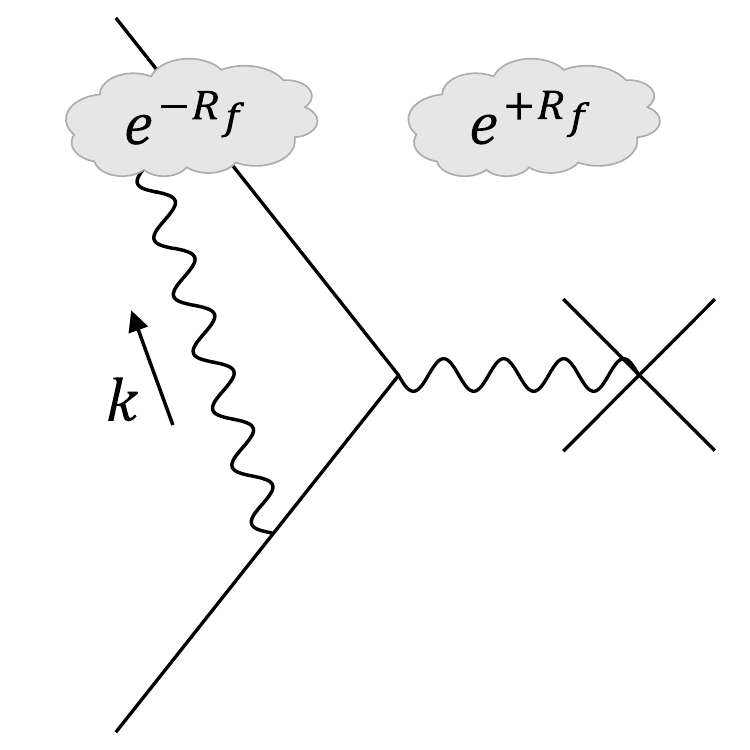}
		\caption{}
    \end{subfigure}
    \begin{subfigure}{0.24\textwidth}
		\includegraphics[width=\textwidth]{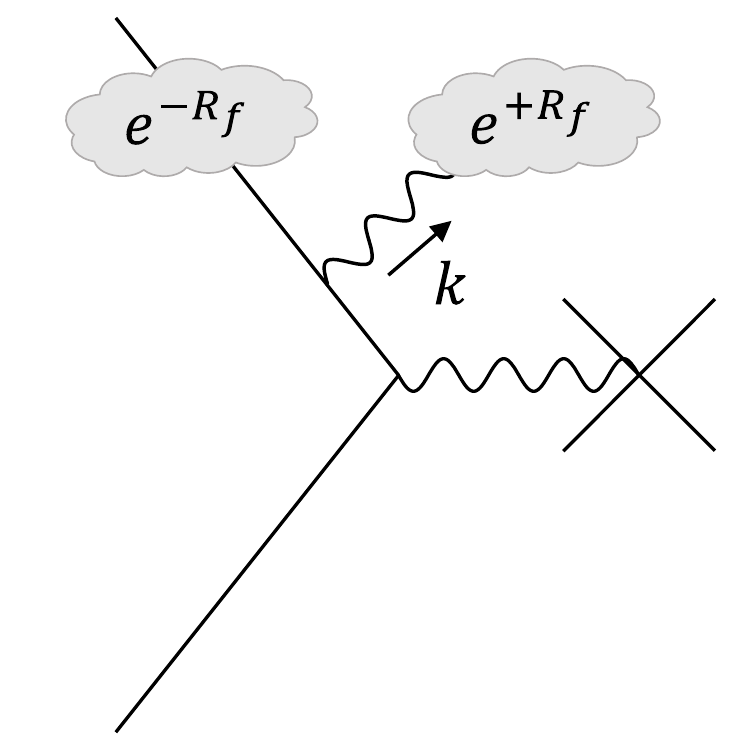}
		\caption{}
    \end{subfigure}
    \begin{subfigure}{0.24\textwidth}
		\includegraphics[width=\textwidth]{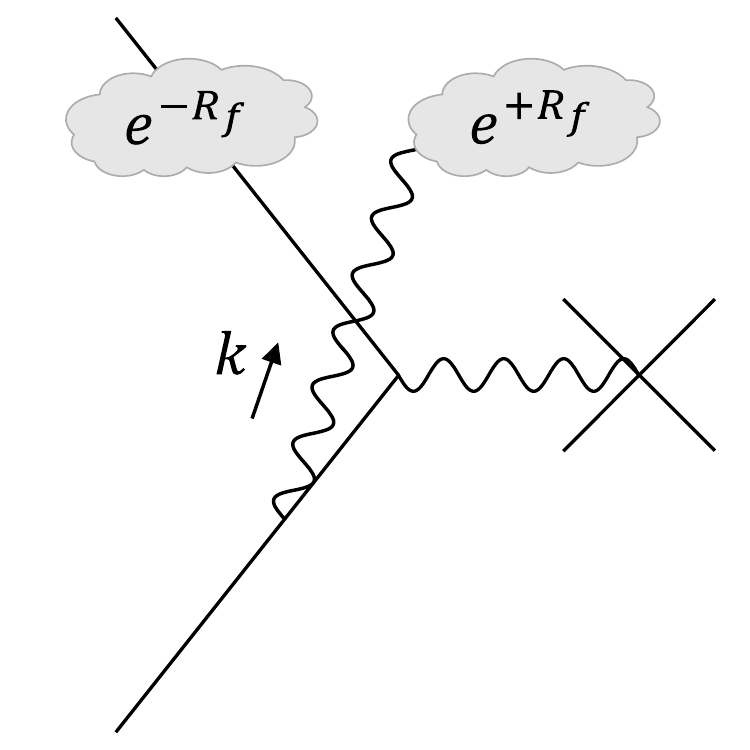}
		\caption{}
    \end{subfigure}
    \caption{Diagrams (a)-(d) represent processes with Faddeev-Kulish asymptotic states.
    	Diagrams (e)-(h) represent the same processes with the incoming cloud moved
    	to the outgoing state. Notice the ``wrong" sign $+R_f$ compared to a normal outgoing cloud with $-R_f$.}
    \label{movecloud}
\end{figure}
Now let us see what happens if we put all the clouds in the outgoing state.
We will denote this amplitude as,
\begin{align}
	\mM_c = 
		\bra{0}
		b(p_f)
		e^{
			-S_f \cdot(a^\dagger - a)
		}
		e^{
			S_i \cdot(a^\dagger - a)
		}
		\mS
		b^\dagger(p_i)
		\ket{0}.
\end{align}
Let us consider the various infrared divergent contributions in this case. The virtual graviton contribution remains unchanged.
For the interacting gravitons, it used to be that the graviton contractions with a cloud gives the factor
\begin{align}
	\frac{1}{2}\int\td{k}S^\mn I_\mnrs,
\end{align}
and depending on whether it was an incoming or an outgoing cloud, the contraction became
\begin{align}
	+\frac{\eta}{2}\int\td{k}S^\mn_f I_\mnrs P^\rs \quad&\text{for outgoing cloud (Figures \ref{movecloud}(a),(b)), and}\\
	-\frac{\eta}{2}\int\td{k}S^\mn_i I_\mnrs P^\rs \quad&\text{for incoming cloud (Figures \ref{movecloud}(c),(d)),}
\end{align}
due to the difference in the sign of soft factor for absorption and emission.
Figures \ref{movecloud}(a) and \ref{movecloud}(c) have $\eta=+1$, while \ref{movecloud}(b) and \ref{movecloud}(d)
	have $\eta=-1$.
But now, we have two clouds that are in the outgoing state, so the graviton contraction gives the factor
\begin{align}
	+\frac{1}{2}\int\td{k}S_f^\mn I_\mnrs \quad&\text{for the $p_f$ cloud, and}\\
	-\frac{1}{2}\int\td{k}S_i^\mn I_\mnrs \quad&\text{for the $p_i$ cloud,}
\end{align}
due to the difference in the signs of $R_f$.
Since both are outgoing clouds, we have the same sign for the soft factor,
\begin{align}
	+\frac{\eta}{2}\int\td{k}S^\mn_f I_\mnrs P^\rs \quad&\text{for the $p_f$ cloud (Figures \ref{movecloud}(e),(f)), and}\\
	-\frac{\eta}{2}\int\td{k}S^\mn_i I_\mnrs P^\rs \quad&\text{for the $p_i$ cloud (Figures \ref{movecloud}(g),(h)),}
\end{align}
where Figures \ref{movecloud}(e) and \ref{movecloud}(g) have $\eta=+1$, while \ref{movecloud}(f) and \ref{movecloud}(h)
	have $\eta=-1$.
One can see that the results stay the same, meaning that contributions of interacting gravitons are unaltered.
It remains to check the cloud-to-cloud contributions, but since these arise from
	contractions between operators in the clouds, they do not depend on which
	side of $\mS$ the cloud is located and therefore are unchanged.
We have thus shown that the infrared divergent part of the single-particle, leading order amplitudes  $\mM_c $ and $\mM_\text{FK}$ remains
	unchanged upon shifting the cloud around, i.e. between the in and out states.

Next, we will generalize this result to the most general case of multiple external particles and all loop orders.
Again, we begin by considering the individual contributions, i.e., virtual, interacting, and cloud-to-cloud gravitons.
The virtual graviton contribution is unchanged from the one given in \cite{Choi:2017bna}.
For the interacting gravitons, consider the amplitude of a diagram with $N$ ($N'$)
	absorbed (emitted) interacting gravitons,
\begin{align}\label{multint}
\begin{split}
	(-1)^N&
	\left[
		\prod_{r=1}^{N+N'}\frac{1}{2}\int\td{k_r}S_\mn(p_r,k_r) I^{\mn \rho_r \sigma_r}
	\right]
	\mJ_{\rho_1\sigma_1\rho_2\sigma_2\cdots\rho_{N+N'}\sigma_{N+N'}},
\end{split}
\end{align}
where $p_r$ is the momentum of the external particle that exchanges graviton $r$,
	and $\mJ$ is a complicated tensor whose detailed form is given in equation (B.58) of \cite{Choi:2017bna}.
Taking the $k$-th incoming cloud and moving it to the outgoing state will have the two following effects (see Figure \ref{factors}):
\begin{figure}
	\centering
	\includegraphics[width=.95\textwidth]{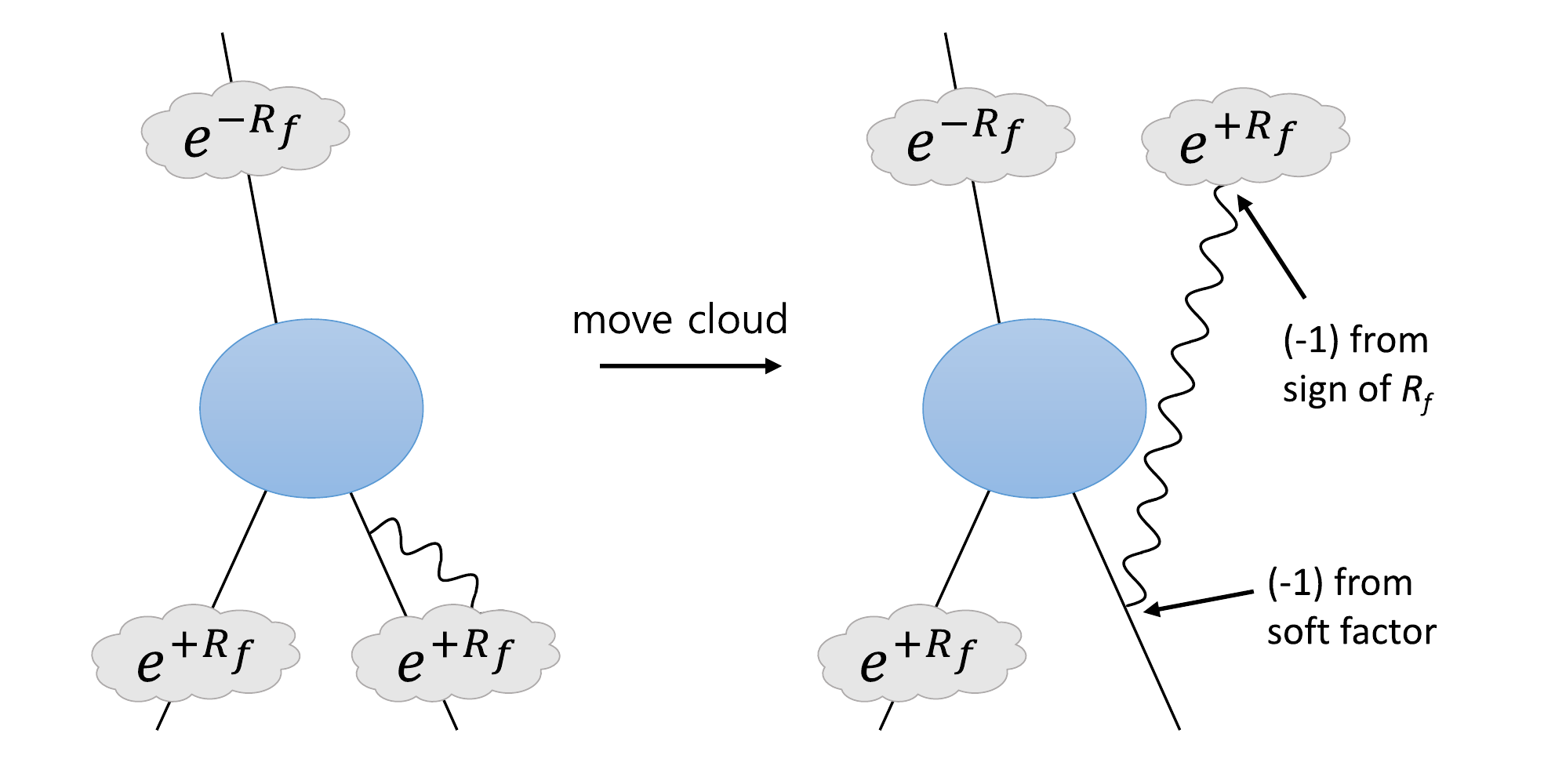}
    \caption{An example of an incoming cloud being moved to the out-state.
    	Each boson connecting this cloud to an external propagator obtains two factors of $(-1)$,
    	one from the soft factor and the other from the ``wrong" sign of $R_f$.
    	These two factors cancel, and thus the overall amplitude is unaffected by such a change.}
    \label{factors}
\end{figure}
\begin{enumerate}
\item The factor $(-1)^N$, which comes from the signs in the soft factors, will become $(-1)^{N-n_k}$,
	where $n_k$ is the number of interacting gravitons connected to the $k$-th (previously) incoming cloud.
	This is because these gravitons used to be absorbed but are now emitted.
\item The following factor in \eqref{multint},
	\begin{align}
		\left[
			\prod_{r=1}^{N+N'}\frac{1}{2}\int\td{k_r}S_\mn(p_r,k_r) I^{\mn\rho_r\sigma_r}
		\right],
	\end{align}
	came from contractions of gravitons with the clouds.
	For the Faddeev-Kulish amplitude, where all clouds are in the proper locations,
		each cloud gives the same factor $\frac{1}{2}\int\td{k}S^\mn I_\mnrs$
		upon contraction.
	But now that the $k$-th incoming cloud is sitting in the outgoing state with a wrong sign
		(incoming and outgoing clouds have different signs $e^{\pm R_f}$), only this cloud gives
		an additional factor of $-1$.
	The above factor changes to
	\begin{align}
		\left[
			(-1)^{n_k}\prod_{r=1}^{N+N'}\frac{1}{2}\int\td{k_r}S_\mn(p_r,k_r) I^{\mn\rho_r\sigma_r}
		\right].
	\end{align}
\end{enumerate}
It follows that we obtain two factors $(-1)^{-n_k}$ and $(-1)^{n_k}$,
	which cancel each other and the overall contribution remains unchanged.
It remains to consider the cloud-to-cloud gravitons.
There are three types: out-to-out, in-to-in, and the disconnected.
The contributions of $l$ disconnected gravitons factored out as
\begin{align}
l!\left[\frac{1}{2}
	S^\text{out}\cdot I\cdot S^\text{in}\right]^l,
\end{align}
but with the $k$-th incoming cloud moved to the out-state (as an outgoing cloud with the wrong sign), this is adjusted to
\begin{align}
l!\left[\frac{1}{2}
	(S^\text{out}-S^k)\cdot I\cdot (S^\text{in} - S^k)
	\right]^l,
\end{align}
which eventually exponentiates to
\begin{align}\label{cloud1}
	\exp\left\{
		\frac{1}{2}(S^\text{out}-S^k)\cdot I\cdot (S^\text{in} - S^k)
	\right\}.
\end{align}
The in-to-in and out-to-out contributions change from
\begin{align}
	\exp\left\{
		-\frac{1}{4}S^\text{out}\cdot I\cdot S^\text{out}
		-\frac{1}{4}S^\text{in}\cdot I\cdot S^\text{in}
	\right\}
\end{align}
to
\begin{align}\label{cloud2}
	\exp\left\{
		-\frac{1}{4}(S^\text{out}-S^k)\cdot I\cdot (S^\text{out}-S^k)
		-\frac{1}{4}(S^\text{in}-S^k)\cdot I\cdot (S^\text{in}-S^k)
	\right\}.
\end{align}
Putting \eqref{cloud1} and \eqref{cloud2} together, we obtain
\begin{align}
	\exp\left\{
		-\frac{1}{4}(S^\text{out}-S^\text{in})\cdot I \cdot (S^\text{out}-S^\text{in})
	\right\},
\end{align}
which is the same factor that was obtained without moving the cloud, and thus
	the cloud-to-cloud contribution also remains unaltered.

It follows that we can write
\begin{align}
	&\bra{\fout}
	\left[\prod_{j\in\text{out}}e^{-S_j\cdot (a^\dagger-a)}\right]
	\mS
	\left[\prod_{i\in\text{in}}e^{S_i\cdot (a^\dagger-a)}\right]
	\ket{\fin}
	\\ &=
	\bra{\fout}
	\left[\prod_{j\in\text{out}}e^{-S_j\cdot (a^\dagger-a)}\right]
	\left[\prod_{i\in\text{in}}e^{S_i\cdot (a^\dagger-a)}\right]
	\mS
	\ket{\fin}
	\\ &=
	\bra{\fout}
	\mS
	\left[\prod_{j\in\text{out}}e^{-S_j\cdot (a^\dagger-a)}\right]
	\left[\prod_{i\in\text{in}}e^{S_i\cdot (a^\dagger-a)}\right]
	\ket{\fin},
\end{align}
and so on.
Therefore, we conclude that the Faddeev-Kulish amplitude does not change under a shift of the cloud from one side of
	the scattering operator to the other.

\subsection{Equality of the amplitudes}\label{SAME}

From \eqref{FKamp} and \eqref{new_amp}, it is clear that the only difference between
	$\mM_c$ and $\mM_\text{FK}$ is in the location of the clouds;
	the incoming cloud, which should be dressing the incoming
	state, is located in the out-state.
We have seen in the previous subsection that in an amplitude the clouds can freely
	be commuted through the scattering operator.
This implies that the amplitude $\mM_c$, which has all the clouds in the outgoing state,
	is actually equal to the Faddeev-Kulish amplitude, i.e.
\begin{align}
	\mM_c = \mM_\text{FK}.
\end{align}
Now let us consider the original amplitude $\mM$ of \eqref{new_amp} that emerged from the
	conservation of supertranslation charge.
This is a special case of $\mM_c$, in the sense that putting $c^\mn = 0$
	in $\mM_c$ recovers $\mM$.
Thus, $\mM$ is equal to the Faddeev-Kulish amplitude
	constructed using the $R(t)$ operator of \cite{Ware:2013zja} instead of $R_f$.
Since states constructed with $R(t)$ and $R_f$ are related by a unitary transformation,
	this implies that $\mM = \mM_c$.
We can see this more directly by noting that amplitudes constructed with $c^\mn=0$
	are related to those with non-zero $c^\mn$ by the following relation \cite{Choi:2017bna}
\begin{align}
	\mM_c &= \exp\left[-\frac{\kappa^2}{4}\sum_{n,m}\eta_n\eta_m
		\int\frac{\td{k}}{\omega_\V{k}^2} c_\mn(p_n,k)I^\mnrs c_\rs(p_m,k)\right]\mM
		= \mM,
\end{align}
where each sum runs over the whole set of external particles.
The summand vanishes term by term, due to one of the constraints that $c^\mn$ has to satisfy.
Therefore $\mM = \mM_c = \mM_\text{FK}$, and the amplitude $\mM$ of \eqref{new_amp}
	is the IR-finite Faddeev-Kulish amplitude.

\section{Soft gravitons and decoherence of momentum configurations of hard matter particles}\label{DECOHERENCE}
In this section we will reconsider the problem of the decoherence of momentum superpositions of hard matter particles due to low energy soft gravitons that was discussed in 
\cite{Carney:2017jut}. The same conclusions were reached in \cite{Strominger:2017aeh} using a different approach.
In \cite{Carney:2017jut} the usual Bloch-Nordsieck mechanism was introduced to cancel the infrared divergences in order to obtain finite density matrices, and the asymptotic symmetries discussed in section \ref{EIGENSTATE} do not play any role. The question we address in this section is how a consistent application of the results of the previous sections might change the conclusions of \cite{Carney:2017jut}. 

First, we will briefly outline the logic of \cite{Carney:2017jut}. 
Consider an ``in" Fock state $\ket{\alpha}_\tin$ at time $t=-\infty$, which is related to the ``out" Fock state at $t=\infty$ by the S matrix:
\begin{align}
	\ket{\alpha} \quad\to\quad \ket{\alpha}_\tin&=\mS \ket{\alpha}_\tout
	\\ &= \left(\sum_{\beta b}\ket{\beta b}\bra{\beta b}\right)
		\mS \ket{\alpha}_\tout
	\\ &= \sum_{\beta b}S_{\beta b,\alpha}\ket{\beta b}_\tout,
\end{align}
where, $S_{\beta b,\alpha} \equiv \braket{\beta b|\mS|\alpha}$, and $\beta$ ($b$) stands for the set of hard (soft) particles.
We will drop subscripts on the kets which, unless specified, will always be the asymptotic out-states.
Then the authors construct a reduced density matrix by tracing out the external
	soft bosons $\ket{b}$:
\begin{align}\label{reduced_dm}
	\rho = \sum_{\beta\beta' b}S_{\beta b,\alpha}S^*_{\beta' b,\alpha}\ket{\beta}\bra{\beta'}.
\end{align}
By factoring out the divergences from the sum, 
\begin{align*}\label{oldform}
	\sum_b S_{\beta b,\alpha}S^*_{\beta' b,\alpha}
		&= S_{\beta,\alpha}S^*_{\beta',\alpha}
			\underbrace{
				\left(\frac{E}{\lambda}\right)^{\tilde{A}_{\beta\beta',\alpha}}
				\left(\frac{E}{\lambda}\right)^{\tilde{B}_{\beta\beta',\alpha}}
				f\left(\frac{E}{E_T},\tilde{A}_{\beta\beta',\alpha}\right)
				f\left(\frac{E}{E_T},\tilde{B}_{\beta\beta',\alpha}\right)
			}_\text{real soft bosons}
		\\
		&= S^\Lambda_{\beta,\alpha}S^{\Lambda*}_{\beta',\alpha}
			\underbrace{
				\left(\frac{\lambda}{\Lambda}\right)^{A_{\beta,\alpha}/2+A_{\beta',\alpha}/2}
				\left(\frac{\lambda}{\Lambda}\right)^{B_{\beta,\alpha}/2+B_{\beta',\alpha}/2}
			}_\text{virtual bosons}
			\\ &\qquad\times
			\underbrace{
				\left(\frac{E}{\lambda}\right)^{\tilde{A}_{\beta\beta',\alpha}}
				\left(\frac{E}{\lambda}\right)^{\tilde{B}_{\beta\beta',\alpha}}
				f\left(\frac{E}{E_T},\tilde{A}_{\beta\beta',\alpha}\right)
				f\left(\frac{E}{E_T},\tilde{B}_{\beta\beta',\alpha}\right)
			}_\text{real soft bosons},
\end{align*}
and by considering the limit as the IR cut-off $\lambda$ is removed, the authors of \cite{Carney:2017jut} observed the decoherence of momentum configurations of hard particles or conversely,  the strong correlations between the hard and soft particles. We will refer to \cite{Carney:2017jut} for details of the notations and derivations of this equation. However, note that it is essential in this approach to sum over the soft bosons because otherwise the infrared divergences will not
cancel.

We now show that this conclusion implicitly assumes that the vacuum is unique and before the cancellation of IR divergences for the inclusive process,  one is dealing with S matrix elements which vanish as the cut-off is removed.
We have seen that conservation of BMS charge, namely
\begin{align}
	\left(N_\tout - N_\tin\right)\braket{\tout|\mS|\tin} = \Omega^\text{soft}\braket{\tout|\mS|\tin},
\end{align}
dictates that scattering processes starting from a state built on the Fock vacuum $\ket{0}$
	evolves only into states that are built on the coherent vacuum
\begin{align}
	\exp\left[\int_\text{soft}\td{k}N_\tout^\mn (a^\dagger_\mn - a_\mn)\right] \ket{0},
\end{align}
where,
\begin{align}
	N_\tout^\mn = \frac{\kappa}{2}\sum_{i}\eta_i\frac{p_i^\mu p_i^\nu}{p_i\cdot k},
\end{align}
with the sum running over all external particles\footnote{Note that here the in and out labels refer to incoming or outgoing particles. The Fock states are all in the ``out" basis.}.
Therefore, if we started with a state $\ket{\alpha}$ built on $\ket{0}$, then the outgoing state cannot be
	just $\ket{\beta b}$, which is a Fock state built on $\ket{0}$; all S-matrix elements between such states will vanish.
Instead, $\ket{\alpha}$ will scatter into states accompanied by a coherent cloud,
\begin{align}
	\ket{\beta;N_\tout}=\ket{\beta}\exp\left[\int_\text{soft}\td{k}N_\tout^\mn (a^\dagger_\mn - a_\mn)\right],
\end{align}
with $N_\tout$ dependent on the sets of external hard momenta $\alpha$ and $\beta$.
We therefore should consider,
\begin{align}
	\ket{\alpha}_\tin
		&= \sum_{\beta} S^\text{FK}_{\beta,\alpha}\ket{\beta;N_\tout},
\end{align}
where we have written,
\begin{align}
	S_{\beta,\alpha}^\text{FK}
		\equiv
			\bra{\beta}
			\exp\left[-\int_\text{soft}\td{k}N^\mn_\tout (a^\dagger_\mn-a_\mn)\right]
			\mS
			\ket{\alpha}.
\end{align}
The states $\ket{\alpha}$ and $\ket{\beta}$ are just the conventional Fock states.
We have seen earlier that the right hand side is exactly equivalent to the amplitude constructed using the Faddeev-Kulish asymptotic states, i.e.,
\begin{align}
	\bra{\beta}e^{-R_f}\mS e^{R_f}\ket{\alpha},
\end{align}
and hence the the left hand side has the superscript $\text{FK}$ on the S matrix element.
Now the density matrix becomes
\begin{align}\label{newrho}
	\sum_{\beta \beta'}S^{\text{FK}}_{\beta,\alpha}S^{\text{FK}*}_{\beta',\alpha}
		\ket{\beta;N_\tout}\bra{\beta';N_\tout'}.
\end{align}
The amplitudes $S^{\text{FK}}_{\beta,\alpha}$ do not have infrared divergences coming from the
	virtual bosons.
In the ``virtual bosons" part of \eqref{oldform}, the $\lambda$-dependent part is exactly canceled by
	interactions involving the clouds, as seen in \cite{Choi:2017bna}. 
Thus, in this framework there is no longer the decoherence that  was observed in \cite{Carney:2017jut}.

To sum up, due to the conservation of BMS charge, any conventional Fock state $\ket{\alpha}$ evolves not into another Fock
	state $\ket{\beta b}$, but instead into a coherent state $\ket{\beta;N_\tout}$.
If the starting state is a coherent state, then the end state will just be another coherent state,
	and the BMS charge conservation will guarantee that the amplitudes $S^\text{FK}_{\beta,\alpha}$ coincide with
	the infrared-finite Faddeev-Kulish amplitudes. We reiterate, that
the presence of the coherent boson cloud cancels all the problematic dependence on the infrared cut-off $\lambda$,
	and therefore one is no longer mathematically forced to sum over the soft particles in order to obtain
	well-defined density matrix elements.

It is noteworthy that although the density matrix elements \eqref{newrho}
	are now well-defined, depending on what kind of measurement is being carried out, one may still construct a reduced density matrix by summing over
	the soft particles. Would the decoherence of the momentum configurations of the hard matter particles return in this case?
This analysis has recently been carried out in \cite{Carney:2017oxp}. We will next reanalyze this within the framework introduced in the previous sections of this paper.

The $\beta \beta'$-component of the reduced density matrix is
\begin{align}
	\rho_{\beta\beta'}
		&= \sum_b
			S^{\text{FK}}_{\beta,\alpha}S^{\text{FK}*}_{\beta',\alpha}
			\braket{b|N_\tout}\braket{N_\tout'|b}
		\\&= 
			S^{\text{FK}}_{\beta,\alpha}S^{\text{FK}*}_{\beta',\alpha}
			\bra{N_\tout'}\left(\sum_b\ket{b}\bra{b}\right)\ket{N_\tout}
		\\&=
			S^{\text{FK}}_{\beta,\alpha}S^{\text{FK}*}_{\beta',\alpha}
			\braket{N_\tout'|N_\tout}.
\end{align}
By normal-ordering the graviton operators, we obtain
\begin{align}
	\braket{N_\tout'|N_\tout}
	&= \bra{0}\exp\left \{
		\frac{\kappa}{2}\int_\text{soft}\td{k}
			\left(
				N_\tout^\mn - N'^\mn_\tout
			\right)
		(a_\mn^\dagger - a_\mn)
	\right \}\ket{0}
	\\ \label{vanishing_W2}
	&=
		\exp\left \{
			-\frac{\kappa^2}{16}\int_\text{soft}\td{k}
			\left(
				N_\tout^\mn - N'^\mn_\tout
			\right)
			I_\mnrs
			\left(
				N_\tout^\rs - N'^\rs_\tout
			\right)
		\right \},
\end{align}
where we can write
\begin{align}
	N^\mn_\tout - N'^\mn_\tout = \sum_{p\in\beta}\frac{p^\mu p^\nu}{p\cdot k} - \sum_{p\in\beta'}\frac{p^\mu p^\nu}{p\cdot k}.
\end{align}
Therefore, if $\beta \neq \beta'$ then the integral in \eqref{vanishing_W2} is infrared-divergent and
	the expression \eqref{vanishing_W2} vanishes.
This implies that the off-diagonal elements of the reduced density matrix is zero
	and the decoherence of momentum configurations of the hard particles reappears.

Does this conclusion change if we include external states with soft gravitons?
The density matrix with external soft gravitons is
\begin{align}
	\sum_{\beta\beta'bb'}S^\text{FK}_{\beta b,\alpha}S^{\text{FK}*}_{\beta' b',\alpha}
		\ket{\beta b;N_\tout}\bra{\beta' b';N'_\tout},
\end{align}
and the reduced density matrix, after tracing out the soft particles, becomes
\begin{align}
	\rho_{\beta\beta'}
		&=
		\sum_{b''}\sum_{b b'}
		S^\text{FK}_{\beta b,\alpha}S^{\text{FK}*}_{\beta' b',\alpha}
		\braket{b''|b;N_\tout}\braket{ b';N'_\tout|b''}
	\\ &=
		\sum_{b b'}
		S^\text{FK}_{\beta b,\alpha}S^{\text{FK}*}_{\beta' b',\alpha}
		\braket{b';N'_\tout| b;N_\tout}.
\end{align}
Let us employ a notation similar to that of \cite{Carney:2017oxp}:
\begin{align}
	W(\beta) &=
		\exp\left\{
			\frac{\kappa}{2}\int_\text{soft}\td{k}
			\sum_{p\in\beta}\frac{p^\mu p^\nu}{p\cdot k}(a^\dagger_\mn-a_\mn)
		\right\},
	\\
	W^\dagger(\beta') &=
		\exp\left\{
			-\frac{\kappa}{2}\int_\text{soft}\td{k}
			\sum_{p\in\beta'}\frac{p^\mu p^\nu}{p\cdot k}(a^\dagger_\mn-a_\mn)
		\right\},
\end{align}
such that $\ket{b;N_\tout}=W(\beta)\ket{b}$.
Then, the reduced density matrix element is
\begin{align}\label{rdm_ext_soft}
	\rho_{\beta\beta'}=
		\sum_{b b'}
		S^\text{FK}_{\beta b,\alpha}S^{\text{FK}*}_{\beta' b',\alpha}
		\bra{b'}\WW
		\ket{b}.
\end{align}
Let us see what we can say about $\braket{b'|\WW|b}$.
Let $m$ and $n$ be the particle number of $b'$ and $b$, respectively.
Then,
\begin{align}
	\braket{b'|\WW|b} = &
	\bra{0}
	a_{\ell'_1}(k'_1)\cdots a_{\ell'_m}(k'_m)
	W^\dagger(\beta')W(\beta)
	a^\dagger_{\ell_1}(k_1)\cdots a^\dagger_{\ell_n}(k_n)
	\ket{0},
\end{align}
where $\ell'_i$ and $k'_i$ ($\ell_i$ and $k_i$) are the polarization and momentum of the $i$-th graviton in $b'$ ($b$). 
Let us use the shorthand
\begin{align}
W^2 \equiv 	\WW,
\end{align}
and observe that since
\begin{align}
	a_\ell(k) &= \epsilon^{\mn}_\ell(k) a_\mn(k),
	\\
	a^\dagger_\ell(k) &= \epsilon^{\mn*}_\ell(k) a^\dagger_\mn(k),
\end{align}
we have the commutators
\begin{align}
	\left[W^2,a^\dagger_{\ell}(k)\right]
	&= -\frac{\kappa}{2}\int_\text{soft}\td{k'}\sum_{p\in\beta,\beta'}\eta_p\frac{p^\mu p^\nu}{p\cdot k'}
		\left[a^\dagger_\mn(k')-a_\mn(k'),a^\dagger_\ell(k)\right]
		W^2
	\\ &=
		+\frac{\kappa}{2}\sum_{p\in\beta,\beta'}\eta_p\frac{p^\mu p^\nu}{p\cdot k}
		\epsilon^{\ell*}_\mn(k)
		W^2,
	\\
	\left[a_{\ell}(k),W^2\right]
	&= -\frac{\kappa}{2}\int_\text{soft}\td{k'}\sum_{p\in\beta,\beta'}\eta_p\frac{p^\mu p^\nu}{p\cdot k'}
		\left[a_\ell(k),a^\dagger_\mn(k')-a_\mn(k')\right]
		W^2
	\\ &=
		-\frac{\kappa}{2}\sum_{p\in\beta,\beta'}\eta_p\frac{p^\mu p^\nu}{p\cdot k}
		\epsilon^{\ell}_\mn(k)
		W^2,
\end{align}
where $\eta_p=+1$ if $p\in\beta'$ and $\eta_p=-1$ if $p\in\beta$.
Using this, we can commute the left-most creation operator $a^\dagger_{\ell_1}(k_1)$ to the left side of $W^2$
	to obtain
\begin{align}
	\nonumber
	\braket{b'|W^2|b} &=
	\left[\frac{\kappa}{2}
		\sum_{p\in\beta,\beta'}\eta_p\frac{p^\mu p^\nu}{p\cdot k_1}
		\epsilon^{\ell_1*}_\mn(k_1)
	\right]
	\bra{0}
	a_{\ell'_1}(k'_1)\cdots a_{\ell'_m}(k'_m)
	W^2
	a^\dagger_{\ell_2}(k_2)\cdots a^\dagger_{\ell_n}(k_n)
	\ket{0}
	\\&\qquad
	+\bra{0}
		a_{\ell'_1}(k'_1)\cdots a_{\ell'_m}(k'_m)
		a^\dagger_{\ell_1}(k_1)
		W^2
		a^\dagger_{\ell_2}(k_2)\cdots a^\dagger_{\ell_n}(k_n)
		\ket{0}.
\end{align}
However, we will next show that the contribution from the second term in the parentheses is vanishingly small.
To see this, one may consider commuting $a^\dagger_{\ell_1}(k_1)$ all the way to the left,
	aiming to act it on the vacuum.
This will create one term for each annihilation operator which has a factor of the following form,
\begin{align}
	\nonumber
	&\int_\text{soft}\td{k'_j}\td{k_1}
		S^\text{FK}_{\beta b,\alpha}S^{\text{FK}*}_{\beta' b',\alpha}
		\left[a_{\ell'_j}(k'_j),a^\dagger_{\ell_1}(k_1)\right]
	\\\label{momvol}
	&= \delta_{\ell'_j,\ell_1}
		\int d\Omega'_jd\Omega_1
		\delta^2(\Omega'_j-\Omega_1)
		S^\text{FK}_{\beta b,\alpha}S^{\text{FK}*}_{\beta' b',\alpha}
		\int_\text{soft} \frac{|\V{k}'_j|^2 d|\V{k}'_j|}{(2\pi)^3|\V{k}'_j|},
\end{align}
where we have separated out the radial parts from the momentum integrals.
\begin{figure}[t]
	\centering
	\includegraphics[width=.19\textwidth]{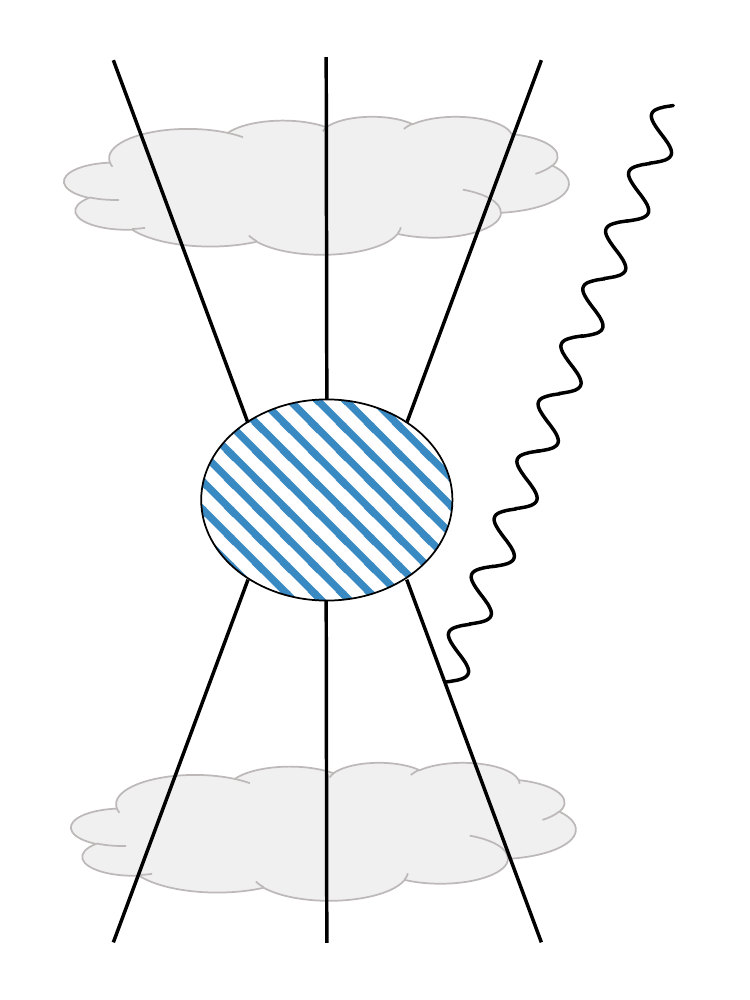}
	\includegraphics[width=.19\textwidth]{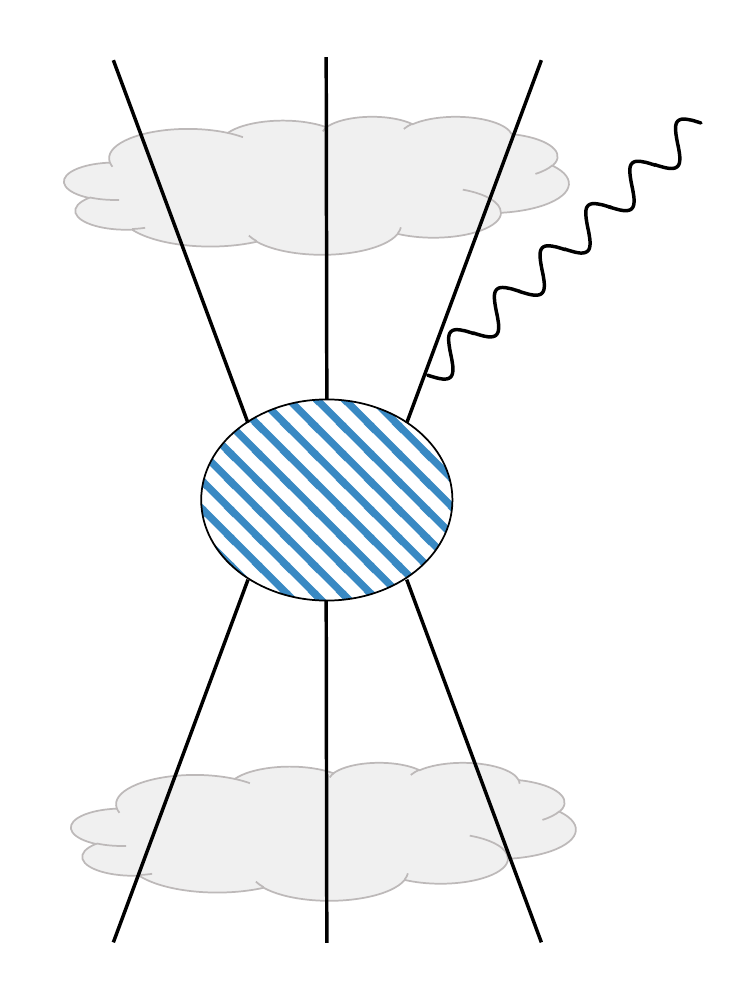}
	\includegraphics[width=.19\textwidth]{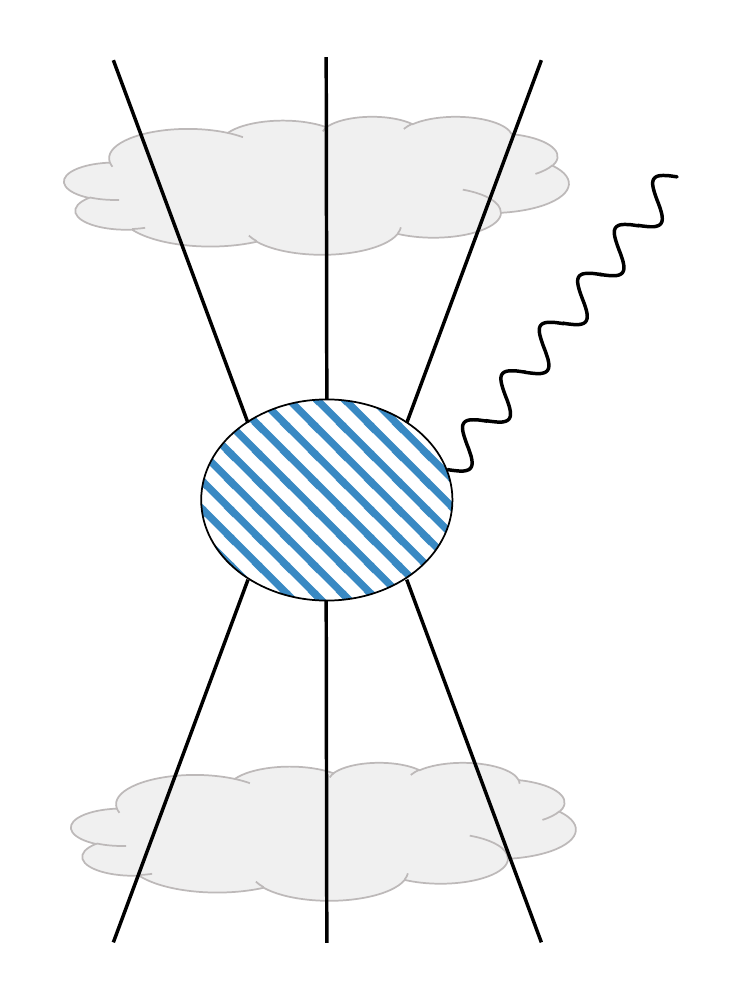}
	\includegraphics[width=.19\textwidth]{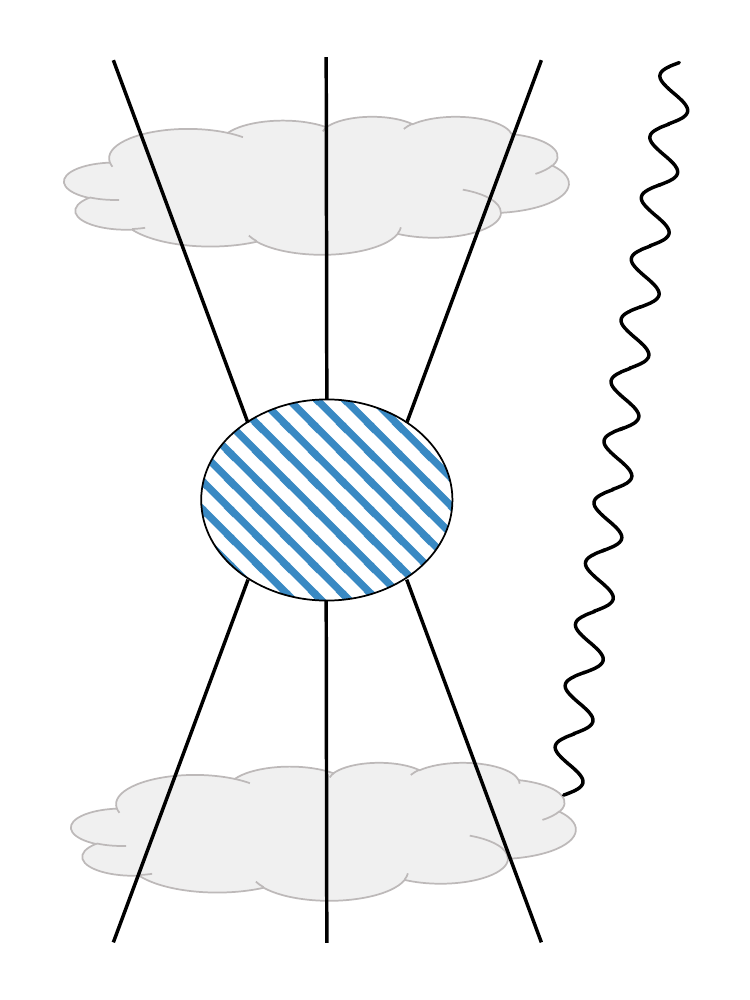}
	\includegraphics[width=.19\textwidth]{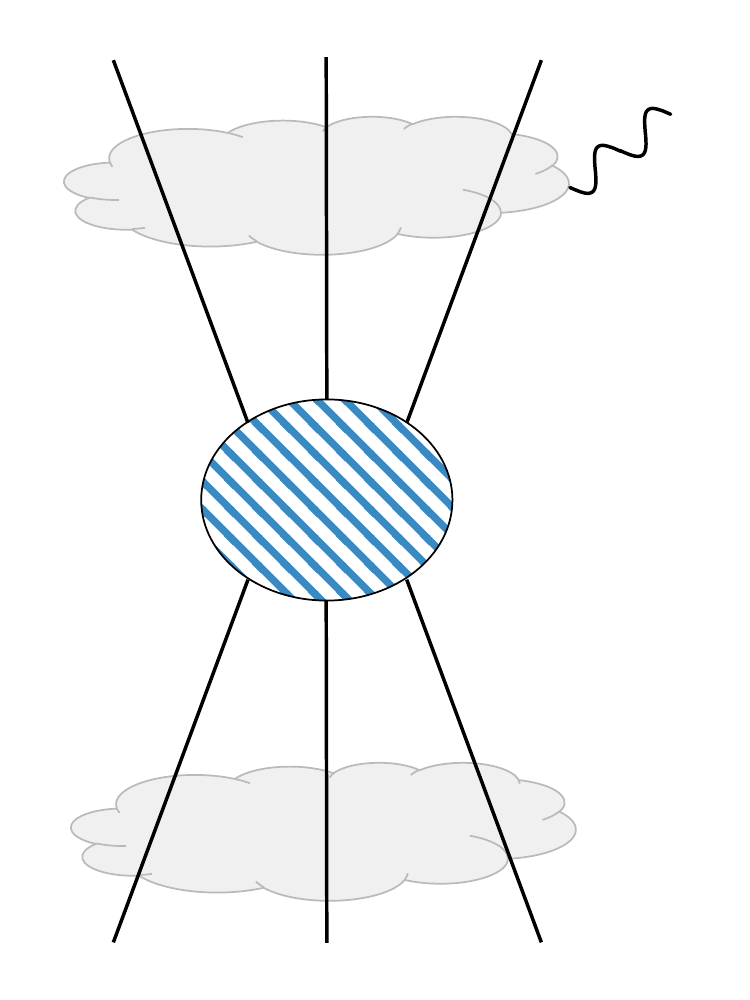}
	\caption{Diagrams contributing to an amplitude with external soft boson.
	The first two diagrams cancel the last two diagrams, and only the diagram in the middle remains, which is
		of zeroth order in the soft momentum.}
	\label{singlegrav}
\end{figure}
The radial integrals can be computed separately,
	because the Faddeev-Kulish amplitudes $S^\text{FK}_{\beta b,\alpha}$
	and $S^{\text{FK}*}_{\beta' b',\alpha}$ are $O(|\V{k}|^0)$ in each soft momentum $\V{k}$,
	which can be seen in figure \ref{singlegrav} for a single outgoing soft graviton;
	the first two diagrams cancel the last two diagrams,
	and only the one in the middle, which is infrared-finite, contribute.
If the momentum integral was over the whole momentum space, then the last integral in \eqref{momvol} will diverge.
But since it is only over the soft momentum space, it has a vanishingly small value
	(proportional to some momentum cutoff squared, $\omega_c^2$, where we think of $\omega_c\to 0$) and therefore the expression vanishes.
Thus, we have
\begin{align}
	\braket{b'|W^2|b} =
	\left[\frac{\kappa}{2}
		\sum_{p\in\beta,\beta'}\eta_p\frac{p^\mu p^\nu}{p\cdot k_1}
		\epsilon^{\ell_1*}_\mn(k_1)
	\right]
	\bra{0}
	a_{\ell'_1}(k'_1)\cdots a_{\ell'_m}(k'_m)
	W^2
	a^\dagger_{\ell_2}(k_2)\cdots a^\dagger_{\ell_n}(k_n)
	\ket{0}.
\end{align}
Each creation operator gives a factor analogous to that in the square brackets, so we may write
\begin{align}
	\braket{b'|W^2|b} =
	\prod_{i=1}^n\left[\frac{\kappa}{2}
		\sum_{p\in\beta,\beta'}\eta_p\frac{p^\mu p^\nu}{p\cdot k_i}
		\epsilon^{\ell_i*}_\mn(k_i)\right]
	\bra{0}
	a_{\ell'_1}(k'_1)\cdots a_{\ell'_m}(k'_m)
	W^2
	\ket{0}.
\end{align}
We can perform a similar process for the annihilation operators, where this time the factors have an additional minus sign,
	and this yields
\begin{align}
	\braket{b'|W^2|b} =
	\prod_{i=1}^n\left[\frac{\kappa}{2}
		\sum_{p\in\beta,\beta'}\eta_p\frac{p^\mu p^\nu}{p\cdot k_i}
		\epsilon^{\ell_i*}_\mn(k_i)\right]
	\prod_{j=1}^m\left[-\frac{\kappa}{2}
		\sum_{p\in\beta,\beta'}\eta_p\frac{p^\mu p^\nu}{p\cdot k'_j}
		\epsilon^{\ell_j}_\mn(k'_j)\right]
	\bra{0}
	W^2
	\ket{0}.
\end{align}
This explicitly shows that each term in the sum of \eqref{rdm_ext_soft} contains a product of infrared-finite integrals
	as well as the vacuum expectation value
	$\braket{0|W^2|0}=\braket{N'_\tout|N_\tout}$,
	but we have seen that this value vanishes for the off-diagonal elements $\beta\neq\beta'$.
Therefore, the reduced density matrix still exhibits a complete decoherence of the hard particle momentum configurations.

We will conclude this section with a discussion of the two formulations of the density matrix: the one using the Bloch-Nordsieck mechanism and the 
one using dressed states. It is straightforward to see that only the off-diagonal elements of the reduced density matrix are different, whereas,
the diagonal element which is essentially the Bloch-Nordsieck cross section is the same in the two approaches. Indeed, the cross section of 
the process $\alpha\to\beta b$ is given (up to a factor) by the absolute square of the
	amplitude:
\begin{align}
	\Gamma_{\beta b,\alpha} = S_{\beta b,\alpha}S^*_{\beta b,\alpha}.
\end{align}
These cross sections exhibit two types of infrared divergence, one arising from the real soft bosons and
	the other from the virtual bosons.
The Bloch-Nordsieck method of dealing with these divergences is to sum over all unobservable soft bosons,
\begin{align}\label{BN}
	\Gamma_{\beta, \alpha} = \sum_b \Gamma_{\beta b,\alpha} = \sum_b S_{\beta b,\alpha}S^*_{\beta b,\alpha},
\end{align}
and performing this sum results in the exponentiation of the soft factors of real bosons, which then cancels the divergence
	due to virtual bosons.
It is clear  that every diagonal element of the reduced density matrix
	in \eqref{reduced_dm} is a Bloch-Nordsieck cross section:
\begin{align}
	\rho_{\beta\beta} = \braket{\beta|\rho|\beta} = \sum_b S_{\beta b,\alpha}S^*_{\beta b,\alpha}.
\end{align}
Thus, only the off-diagonal elements are affected (see Eq. \eqref{rdm_ext_soft}). The practical use of the Bloch-Nordsieck mechanism for obtaining IR finite cross sections does not require any modifications.

\section{Discussion}\label{DISCUSSION}

We have demonstrated that graviton cloud operators weakly commute with the scattering operator,
	and used this to show that scattering amplitudes which conserve BMS supertranslation charge
	are equal to the Faddeev-Kulish amplitudes.
Since Faddeev-Kulish amplitudes are free of infrared divergence, this proves the conjecture in
	\cite{Kapec:2017tkm}, that conservation of asymptotic charge leads to infrared finite
	scattering amplitudes.
Our work ties up some loose ends on the relation between the Faddeev-Kulish formalism,
	asymptotic symmetry and infrared divergences. In particular, it supports the viewpoint that in QED and perturbative gravity, infrared 
	divergences in the usual Dyson expansion of the S matrix are not real in the sense that they arise only as a result of using states that violate the conservation of
	the BMS supertranslation charge. Our paper also clarifies a common misconception in the literature that soft clouds ``surround'' the asymptotic particle. In actual fact, soft photons always go out to null infinity and massive particles to time-like infinity. Thus at large enough retarded times, one has matter particles surrounded by static fields (for example, coulomb fields in QED)\footnote{We thank Andy Strominger for clarifying this.}. Indeed, it was conjectured in \cite{Kapec:2017tkm} and proved in this paper that the soft clouds can be moved from the in state to the out state without loosing infrared finiteness. 

We have also applied our formalism to the intriguing problem considered in \cite{Carney:2017jut} 
where it was found that tracing out soft degrees of freedom leads to the decoherence of hard particle
	momenta, whether or not one employs the Faddeev-Kulish states \cite{Carney:2017oxp}. 
In contrast to their work, we have constructed the corresponding reduced density matrices conserving the BMS supertranslation charge
	at all stages and arrived at a similar conclusion. It seems puzzling how to reconcile this with the fact that 
there are entanglement phenomenon observed in nature. Perhaps the large time S matrix approach may not be well suited to this problem.

It is worth noting that while we have worked exclusively with gravity,
a similar analysis can be applied to  QED. This would lead to an analogous conclusion, namely, 
	that large gauge transformation charge conservation implies infrared-finite Faddeev-Kulish amplitudes of QED.
Therefore, a natural direction for future study would be the extension of these results to QCD.
It will be very interesting and non-trivial to see how the asymptotic symmetry of QCD relates
to the infrared-finite Faddeev-Kulish amplitude in that theory. More specifically, in QCD what is the connection between large gauge transformation
charge conservation and infrared finiteness of the appropriate S matrix elements?

\acknowledgments
S.C. acknowledges a fellowship from the Samsung Foundation of Culture.
We are grateful to Andy Strominger for his insightful comments on the manuscript. We would like to 
thank Sandeep Pradhan for an initial collaboration and him, Gordon Semenoff  and Malcolm Perry for discussions.

\bibliographystyle{jhep} 
\bibliography{references}

\end{document}